\documentclass[reprint, superscriptaddress, amsmath, amssymb, aps, prx, floatfix]{revtex4-2}

\usepackage{braket}
\usepackage{printlen} 
\usepackage{amsthm}
\usepackage{gensymb}
\usepackage{isotope}
\usepackage[
    per-mode=symbol,
    retain-zero-uncertainty=true
    ]{siunitx}
\DeclareSIUnit{\gauss}{G}
\DeclareSIUnit{\phonons}{phonons}
\usepackage{braket}
\usepackage{color}
\usepackage{graphicx}
\graphicspath{{supp_mat_figures/}}
\usepackage{dcolumn}
\usepackage{bm}
\usepackage{hyperref}
\hypersetup{
pdftitle={Two-dimensional shelving spectroscopy of ultraviolet ground state transitions in dysprosium},
pdfauthor={Dysi Team},
pdfkeywords = {dysprosium, polarizability, spectroscopy},
colorlinks=true, linkcolor=[RGB]{0,0,0}, citecolor=magenta, filecolor=[RGB]{0,0,0}, urlcolor=[RGB]{0,0,0}}

\usepackage[
    per-mode=symbol,
    retain-zero-uncertainty=true
    ]{siunitx}

\usepackage{bm}
\usepackage{enumerate}
\usepackage{tikz}
\usepackage{dsfont}
\usetikzlibrary{quantikz}
\usepackage{lipsum} 
\usepackage{subcaption}
\usepackage{ragged2e}
\DeclareCaptionJustification{justified}{\justifying}
\usepackage{printlen}
\usepackage{etoolbox}
\makeatletter
\def\maketitle{
\@author@finish
\title@column\titleblock@produce
\suppressfloats[t]}
\makeatother
\newcounter{PRLsections}

\newcounter{PRLsubsections}[section]

\usepackage{xpatch}
\xpretocmd{\section}{\setcounter{PRLsubsections}{0}}{}{}

\usepackage{wasysym}
\usepackage[OT2,T1]{fontenc}
\DeclareSymbolFont{cyrletters}{OT2}{wncyr}{m}{n}
\DeclareMathSymbol{\Sha}{\mathalpha}{cyrletters}{"58}

\usepackage[innercaption]{sidecap}
\sidecaptionvpos{figure}{c}
\DeclareSymbolFont{yhlargesymbols}{OMX}{yhex}{m}{n} 
\DeclareMathAccent{\widehat}{\mathord}{yhlargesymbols}{"62}
\usepackage{multirow}
\usepackage{array}
\newcolumntype{L}[1]{>{\raggedright\let\newline\\\arraybackslash\hspace{0pt}}m{#1}}
\newcolumntype{C}[1]{>{\centering\let\newline\\\arraybackslash\hspace{0pt}}m{#1}}
\newcolumntype{R}[1]{>{\raggedleft\let\newline\\\arraybackslash\hspace{0pt}}m{#1}}

\usepackage{pgfplots}
\usepackage{pgf}
\usepackage{tabularx} 
\usepackage{booktabs}
\usepackage{threeparttable}
\usepackage{makecell}
\usepackage{coffeestains}

\newcommand{\be}{\begin{equation}} 
\newcommand{\ee}{\end{equation}}

\newtoggle{arXiv} 

\toggletrue{arXiv}      
\begin{document}



\title{Two-dimensional shelving spectroscopy of ultraviolet ground state transitions in dysprosium}

\author{Kevin S. H. Ng}
\email[E-mail: ]{kevin.ng650@gmail.com}
\author{Paul Uerlings}
\author{Fiona Hellstern}
\affiliation{5. Physikalisches Institut and Center for Integrated Quantum Science and Technology, Universität Stuttgart, Pfaffenwaldring 57, 70569 Stuttgart, Germany} 
\author{Jens Hertkorn}
\affiliation{5. Physikalisches Institut and Center for Integrated Quantum Science and Technology, Universität Stuttgart, Pfaffenwaldring 57, 70569 Stuttgart, Germany} 
\affiliation{Department of Physics, MIT-Harvard Center for Ultracold Atoms, and Research Laboratory of Electronics, MIT, Cambridge, Massachusetts 02139, USA}
\author{Luis Weiß}
\author{Stephan Welte}
\author{Tilman Pfau}
\email[E-mail: ]{t.pfau@physik.uni-stuttgart.de}
\author{Ralf Klemt}
\email[E-mail: ]{rklemt@pi5.physik.uni-stuttgart.de}
\affiliation{5. Physikalisches Institut and Center for Integrated Quantum Science and Technology, Universität Stuttgart, Pfaffenwaldring 57, 70569 Stuttgart, Germany} 
 
\begin{abstract}
\medskip

\noindent The open inner-shell electronic structure of lanthanides with large magnetic moments gives rise to a rich spectrum of transitions available for laser cooling, trapping, and coherent control. Despite this, the large number of ultraviolet (UV) transitions below \SI{400}{\nano m} have so far been rarely utilized in dipolar atom experiments. Here, we investigate multiple UV ground state transitions in dysprosium. Several of these UV excited states have the largest decay strengths to the ultralong-lived, low-lying first excited state which are comparable to the most commonly used strongest transitions found in dipolar atoms. Using two-dimensional shelving spectroscopy which improves detection sensitivity and provides a straightforward way to determine the hyperfine-isotope structure and excited state total angular momentum $J$, we measure isotope shifts, hyperfine coefficients, and create King plots to determine their electronic nature. Such knowledge of these UV transitions which analogously exist in other magnetic atoms is important for optically populating the first excited state and can be used towards creating an optical clock, high resolution imaging in quantum gas microscopy, and probing lanthanide nuclei with enhanced Schiff moments in search of physics beyond the standard model.

\noindent 
\end{abstract}

\maketitle 

\section{Introduction}

\newcolumntype{Y}{>{\centering\arraybackslash}X}

Magnetic dipolar atoms are prime candidates for understanding and realizing systems with long-range anisotropic dipole interactions \cite{chomaz2023, böttcher2021, kadau2016, schmitt2016, norcia2021, su2023} in addition to studying physical models with large spin \cite{lepoutre2018, chalopin2020}. Preparing and manipulating these systems with magnetic lanthanides in ultracold atom experiments is facilitated by the large number of electronic transitions that range in natural linewidth from the sub-Hz to MHz level, a consequence of their open inner-shell electronic structure. Furthermore, unique states of opposite parity very close in energy have motivated searches in variations of the fine structure constant \cite{budker1993, vantilburg2015} and are also used to induce large electric dipole moments to further enhance long-range interactions \cite{lepers2018, seifert2025}. 

From the wide choice of possible states in lanthanides, the first excited state (FES) stands out as an attractive resource. In dysprosium (Dy), erbium (Er), holmium (Ho) and thulium (Tm), the FES has the same electronic configuration and parity as the ground state. These properties result in a low-lying, ultra-long lived state that shares a very similar dynamic scalar polarizability as the ground state across a wide range of wavelengths \cite{polarisabilityComment}. Such ultranarrow states ($\approx$ Hz \cite{sukachev2016} - \textmu Hz \cite{kozlov2013}) that are insensitive to the black-body radiation frequency shift have been proposed as promising candidates for optical clocks \cite{li2017, dzuba2014, kozlov2013, li2017Holmium, golovizin2019}. Moreover, they can be used for applications in quantum computing and simulation where coherent state control is key \cite{mishin2025}, or even as a tool in high-resolution imaging, where selective shelving of atoms trapped in optical superlattices at magic wavelengths enables sub-diffraction limited resolution \cite{petersen2020, patschneider2021}. In contrast to the narrow \SI{1001}{\nano m} \cite{petersen2020} and \SI{1299}{\nano m} \cite{patschneider2021} lines in Dy and Er respectively, the FES in Dy, Er and Ho that is importantly more conveniently magic with the ground state \cite{li2017, li2017Holmium} has to our knowledge never been optically accessed and utilized in any cold atom experiments. With the non-standard wavelengths ($>$\SI{1800}{\nano m}) and demanding laser stability required, direct access to the FES is challenging. 

\begin{figure*}[ht]
    \centering
    \includegraphics[]{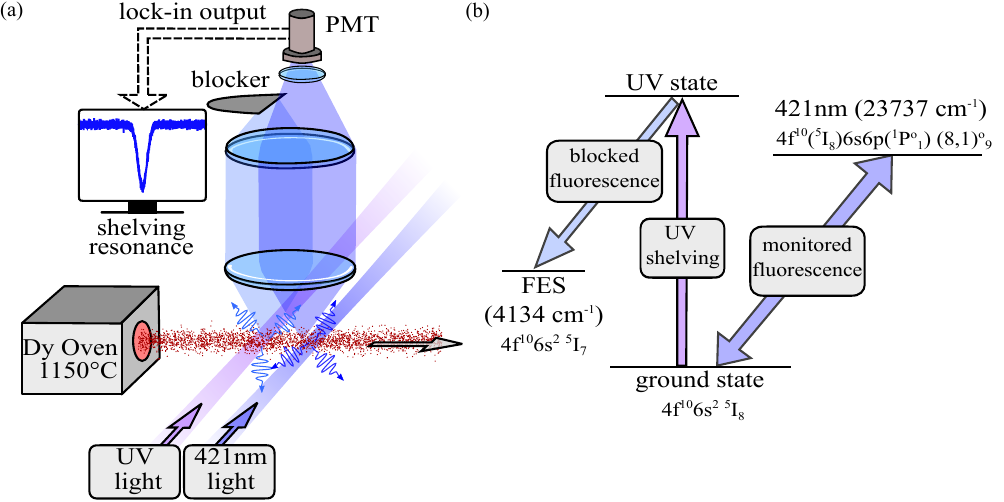}
    \caption{(a) Experimental setup: Dy atoms (red) emerge from an oven and are illuminated subsequently by UV and then \SI{421}{\nano m} laser beams. The \SI{421}{\nano m} beam fluorescence is focused onto the photomultiplier tube (PMT). The blocker blocks any unwanted fluorescence from UV state decay to the FES. (b) Shelving spectroscopy energy level scheme: ground state atoms (electronic configuration [Xe]$4f^{10}6s^{2}$) are shelved by UV light and UV resonances are detected by monitoring the \SI{421}{\nano m} fluorescence. For most of the transitions studied, the decay to the FES at 4134 cm$^{-1}$ is significantly stronger than decay back to the ground state.}
    \label{fig:1}
\end{figure*}

The ground state UV transitions in Dy provide a more accessible pathway to populating the FES. From all possible electric dipole-allowed ground state transitions, several of these transitions have the largest decay rates and branching ratios to the FES \cite{conway1971, wickliffe2000}. The decay rates are comparable to the largest transition rates in Dy and are ideal for populating the FES, both via optical pumping or through fast coherent coupling from the ground state via a two-photon Raman transition \cite{unnikrishnan2024}. In fact, analogous UV transitions exist in the above-mentioned lanthanides \cite{lawler2010, wickliffe1997, alLabady2017} and are also ideal for measuring the isotope shifts of the FES for Dy and Er with high precision via Raman spectroscopy. A King plot analysis of the lowest-lying state provides important information about the shape of Dy and Er nuclei, which are proposed candidates for having enhanced Schiff moments due to their octopole deformation in search of charge-parity violating interactions \cite{engel2025, flambaum2020, flambaum2025}.

In this paper, we measure the isotope shifts and hyperfine structure of UV transitions which lie between 359.0-\SI{372.5}{\nano m}, as well as perform King plot analysis to characterize their electronic nature. Unlike typical single-beam methods employed to measure these values for ground state transitions in lanthanides \cite{leefer2009, lu2011, schmitt2013, patschneider2021, lipert1993, jin2011}, we exploit the shared ground state of the strong blue \SI{421}{\nano m} transition \cite{wickliffe2000} to perform shelving spectroscopy (also known as optical-optical double resonance spectroscopy \cite{demtröder2008, manai2020}). By furthermore varying both the UV and the blue frequencies in our experiment, this two-dimensional shelving spectroscopic technique increases detection sensitivity and simultaneously significantly simplifies the assignment of the numerous isotope and hyperfine transitions that exist due to the large hyperfine state manifold. We show that this technique can be used to determine the total angular momentum $J$ of the excited state without any applied magnetic field, varying light polarization, fitting of measured spectra or prior knowledge of the electronic configuration. Since there are ground state UV transitions in Dy with misassigned values for $J$ in standard spectroscopic tables \cite{white2025}, this technique is useful for extracting $J$ reliably and is advantageous with any transition where dense spectra with many lines are expected. In turn, this further supports more accurate calculations of dynamical polarizabilities \cite{li2017} for optical trapping in short wavelength UV lattices that enhance dipolar interactions in quantum simulations of extended Hubbard models \cite{fraxanet2022}. 

\section{Experimental setup}

\begin{table*}
\caption{Detected UV transitions and the \SI{421}{nm} transition used for shelving detection. Additional previously reported UV transitions between 359.0 and \SI{372.5}{\nano m} are listed in Table \ref{table:S1}. Transitions are referred to using the vacuum wavelength and wavenumber.   For each detected UV state (except \SI{366.08}{\nano m}), only a single side decay channel to the FES at 4134 cm$^{-1}$ has been reported. If no leading term symbol is reported, only $J$ is written.}
\label{table:1}
\begin{threeparttable}
\begin{tabularx}{\textwidth}{Y Y Y c@{}c Y c@{}c} 
 \hline\hline
 \makecell{Wavelength (nm)} & \makecell{Wavenumber (cm$^{-1}$)} & \makecell{Leading term symbol \\ ($^{2S+1}L_{J}$)} & \multicolumn{2}{c}{\makecell{$A_\text{GS}$\\(10$^{6}$ s$^{-1})$ \tnote{a}}} & \makecell{UV-FES decay\\ wavelength (nm)\tnote{d} }& \multicolumn{2}{c}{\makecell{$A_\text{FES}$ \\(10$^{6}$ s$^{-1})$ \tnote{a}}} \\
 \hline
 359.05 & 27851.4247(16)\tnote{c} & $^{7}$I$_{8}$ & 0.326 &\ \ (300) & 421.63 & 81.0 & - \\ 
 359.26 & 27834.9306(16)\tnote{c} & $^{5}$I$_{7}$ & - &\ \ (10) & 421.93 & 120.0 &\ ($10^{4}$)\\ 
 359.48 & 27817.9903(16)\tnote{c} & $^{5}$K$_{8}$ & - &\ \ (100)& 422.23 & 128.0 &\ ($10^{4}$)\\
 362.89 & 27556.3275(16)\tnote{c} & $^{9}$L$_{7}$ & 0.175 &\ \ (300)& 426.95 & 1.14 & \ ($10^{3}$)\\ 
 366.08 & 27316.49\tnote{b} & $^{7}$G$_{7}$ & - & \ \ - & - & - & \ -  \\
 372.19 & 26868.0657(15)\tnote{c} & $J = 8$ & - &\ \ (300)& 439.87 & - &\ ($100$)\\
 421.29 & 23736.60\tnote{d} & $^{5}$K$_{9}$ & 208.0 &\ \ ($10^{4}$) & -& - & \ - \\
 \hline\hline
\end{tabularx}
\begin{tablenotes}
\item[a] values from \cite{wickliffe2000} and values in brackets from \cite{conway1971} given as relative strengths. 
\item[b] from \cite{martin1978}. Detected, however signal to noise ratio was not large enough to extract relevant values.
\item[c] measured absolute wavenumber for the transition for \isotope[164]{Dy} with systematic error (see supp. mat.). 
\item[d] from \cite{conway1971} and \cite{wickliffe2000}.
\end{tablenotes}
\end{threeparttable}
\end{table*}

Our experimental setup and shelving scheme are shown in Figure \ref{fig:1}. An effusion cell containing dysprosium granulate is heated up to $\approx$ 1150 $^{\circ}$C to produce an atomic beam in an ultra-high vacuum ($\approx$10$^{-10}$ mbar) chamber. To reduce the atomic beam spread and resulting residual Doppler broadening in the chamber where we perform our measurements, we install a \SI{22}{\milli m} diameter aperture $\approx$ \SI{11}{\centi m} after the output. This results in a full-angle divergence of $\approx\pi/15$ radians which simultaneously also prevents our vacuum viewports from becoming coated with dysprosium. Orthogonal to the atomic beam, we shine in a single UV beam ($\approx$ \SI{37}{\milli\watt}, $\approx$ \SI{0.47}{\milli m} $1/e^{2}$ waist) and a larger blue beam ($\approx$ \SI{0.85}{\milli\watt}, $\approx$ \SI{0.85}{\milli m} $1/e^{2}$ waist) aligned parallel to each other, $\approx$ \SI{1}{\centi m} apart such that atoms first encounter the UV beam before passing through the blue beam (Figure \ref{fig:1}(a)). At each blue frequency which is subsequently increased in steps of $\approx$ \SI{20}{\mega\hertz}, we then scan the frequency of the UV light across resonance while monitoring the blue beam fluorescence from the strong \SI{421}{\nano m} transition ($\Gamma_{\text{blue}} = 2\pi \cdot32$\,MHz) \cite{lu2011}. When the UV frequency is on resonance, a fraction of the atoms are shelved from the ground state (Figure \ref{fig:1}(b)) and are then no longer excited by the blue beam. The resulting decrease in \SI{421}{\nano m} fluorescence is measured as a shelving resonance that becomes Doppler-free when a fixed velocity class is selected by choosing the relative frequencies accordingly. The shelving scheme relies on some fraction of excited atoms not decaying back to the ground state before they reach the blue beam. In our experiment, this condition is established by a fast decay to the FES at \SI{4134}{\centi m}$^{-1}$ (Table~\ref{table:1}), with a relative decay ratio $A_{\text{FES}}/A_{\text{GS}}$ between $\approx$ 1000-0.3:1 for the transitions studied, where $A_\text{FES}$ and $A_\text{GS}$ are the A coefficients to the first excited state and ground state respectively. By having a \SI{1}{\centi m} beam separation which corresponds to a travel time between beams of $\approx$ \SI{20}{\micro\second} for an average approximate atomic velocity of \SI{500}{ m/\second}, we ensure that decay lifetimes from the \SI{4134}{\centi m}$^{-1}$ state to the ground state are significantly longer than the travel time. With this shelving scheme and a moderate saturation parameter of 1.3 for our blue beam, the detection sensitivity of a single UV excitation is enhanced by a factor of around 180 (see supp. mat.). 

Light generation and detection are performed as follows. We generate UV light with tunable wavelength using a \SI{532}{\nano m}-pumped Ti:Sa crystal (M2 Solstis) which produces light that is frequency-doubled with a lithium triborate crystal (M2 ECD-X) to the desired UV wavelength. The \SI{421}{\nano m} light is produced in a similar way using the \SI{842}{\nano m} light (Coherent MBR 110) frequency-doubled with a home-built cavity. Both systems can produce at least \SI{1}{\watt} of power at the desired wavelengths. The frequency of the \SI{421}{\nano m} light is set at a desired frequency by locking the \SI{842}{\nano m} light to a moveable sideband of a temperature-stabilized ultra-low expansion (ULE) cavity (free spectral range of \SI{1.498}{\giga\hertz}) produced by a fiber-coupled electro-optic modulator. The power is actively stabilized by power modulation of the radio-frequency (RF) supplied to an acousto-optic modulator (AOM). The frequency of the UV light is scanned during spectroscopy by internally locking the doubling cavity while scanning the Ti:Sa frequency, which is simultaneously recorded by a wavemeter (High Finesse WS/8-2). We use a hydrogen-loaded photonic crystal fiber (NKT aeroGuide-PM-10) to prevent fiber solarization due to UV light \cite{colombe2014} to deliver the power to the chamber. 

\begin{figure*}[ht]
    \centering
    \includegraphics[]{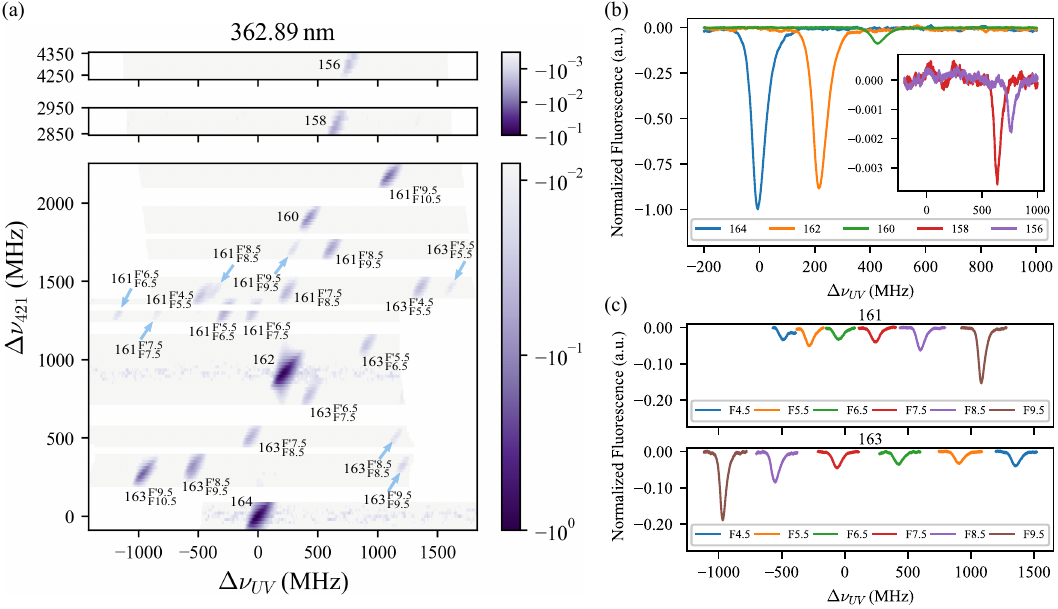}
    \caption{(a) Fluorescence map of the \SI{362.89}{\nano m} transition, normalized to the absolute value of the minimum corresponding to the most abundant isotope \isotope[164]{Dy}. Labels show the involved isotope, subscripts (superscripts) denote the ground (excited) hyperfine state. Blue arrows indicate detection of the less probable $\Delta F = 0$ hyperfine transitions. White spaces indicate frequency intervals where we did not measure. The colorbar indicates the normalized fluorescence (a.u.) with \isotope[158]{Dy} and \isotope[156]{Dy} measured at a different PMT gain setting due to their low abundance. (b) Bosonic and (c) dominant fermionic resonances taken as single traces of (a) with the blue frequency difference $\Delta\nu_{421}$ set to the respective hyperfine transition.}
    \label{fig:2}
\end{figure*}

We collect and focus the monitored blue fluorescence onto a photomultiplier tube (PMT) (Hamamatsu H6780-20) with two lenses in an approximate 4f-configuration and an additional focusing lens (Figure \ref{fig:1}(a)). A linewidth filter centered at \SI{420}{\nano m} (3\,dB width/\SI{10}{\nano m} full-width half maximum) is mounted in front of the PMT and we physically block fluorescence from reaching the PMT produced from decay to the FES at 4134 cm$^{-1}$, as some transitions we study produce fluorescence at wavelengths not attenuated by the filter. The beam separation between UV and blue light is sufficient to block unwanted fluorescence while maintaining a strong 421nm fluorescence signal, and we checked that any unwanted stray fluorescence from decay to the FES is sufficiently suppressed while scanning the UV frequency. The output of the PMT is fed into a lock-in amplifier and referenced with the \SI{3}{\kilo\hertz} modulating signal from a function generator that modulates the RF power driving an AOM that shifts our UV frequency by \SI{110}{\mega\hertz}. The lock-in output was then recorded for all UV transitions that were detectable with our setup to produce spectroscopic maps such as the one in Figure \ref{fig:2}(a).

\section{Experimental results}

\newcolumntype{Y}{>{\centering\arraybackslash}X}

In our experiment, we detect 6 UV transitions from \SI{359.0}{\nano m} to \SI{372.5}{\nano m} \cite{conway1971, ahmad1982, wickliffe2000, wyart1974, martin1978} (Table~\ref{table:1}). 
For 5 of the transitions, we performed shelving spectroscopy to measure the isotope shifts and hyperfine structure of each transition. A shelving spectroscopy map of the measured fluorescence for the \SI{362.89}{\nano m} transition as a function of the UV $\Delta\nu_{\text{UV}}$ and blue $\Delta\nu_{421}$ frequency difference from the extracted transition frequency of \isotope[164]{Dy} is shown in Figure \ref{fig:2}(a). When the frequency of the blue beam is on resonance with a particular isotope or hyperfine transition of the \SI{421}{\nano m} excited state, shelving resonances appear as fluorescence minima when we vary $\Delta\nu_{\text{UV}}$. These resonances correspond to UV shelving of atoms addressing the same isotope or hyperfine ground state as the blue beam. We detect 17 dominant resonances in total (Figure \ref{fig:2}(a)) which exhibit a slope $\Delta\nu_{421}/\Delta\nu_{\text{UV}}$ that closely matches the ratio $k_{421}/k_{\text{UV}}$ $\approx$ 0.86, required for our two wavelengths to address a common Doppler-shifted velocity class. Using the known isotope shifts and hyperfine splitting of the \SI{421}{\nano m} excited state \cite{leefer2009}, the resonances are assigned accordingly to all 5 stable bosonic isotopes of dysprosium (relative abundances between 28\% and 0.06\%), as well as to all 6 possible $\Delta F = F'-F = -1$ excitations between ground and excited hyperfine states for both \isotope[161]{Dy} and \isotope[163]{Dy} ($\Delta F = \Delta J$ are the most probable hyperfine transitions (supp. mat. Table~\ref{table:S2}), where $J = 7$ for this state \cite{wickliffe2000}). Correctly assigning each resonance is straightforward, as resonance signals that would otherwise overlap each other with single-beam spectroscopic methods are well-separated in single traces (Figure \ref{fig:2}(b), (c)). We note a background signal in the fluorescence map due to technical noise from the lock-in amplifier that can be seen at $\Delta\nu_{421} \approx 0$ and $\approx$ \SI{915}{\mega\hertz}, where the largest signals corresponding to \isotope[164]{Dy} and \isotope[162]{Dy} were measured. We do not attribute these small oscillations to any resonances. We check our assignment of all isotopes by comparing their relative strengths seen in the single traces through each resonance (Figure \ref{fig:2}(b), (c)) to the natural isotopic abundance of dysprosium, where we find good agreement.

\begin{table*}
\caption{Measured isotope shifts of each UV transition. Unless otherwise specified, all values are in units of MHz.}
\label{table:2}
\begin{tabularx}{\textwidth}{Y Y Y Y Y Y Y} 
\hline\hline
 \text{Wavelength (nm)} & \text{$\delta\nu_{164-163}$} & \text{$\delta\nu_{164-162}$} & \text{$\delta\nu_{164-161}$} & \text{$\delta\nu_{164-160}$} & \text{$\delta\nu_{164-158}$} & \text{$\delta\nu_{164-156}$} \\  
 \midrule
 359.05 & -256(13) & -447(14) & -730(13) & -921(14) & -1408(14) & -1986(14) \\
 359.26 & -340(14) & -558(14) & -945(14) & -1144(14) & - & -\\
 359.48 & -380(13) & -610(14) & -1034(13) & -1251(14) & -1891(14) & -2757(14) \\
 362.89 & -70(14) & -217(14) & -262(14) & -424(14) & -644(14) & -774(14) \\
 372.19 & -44(10) & -178(13) & -194(13) & -339(13) & - & - \\
 \hline\hline
\end{tabularx}
\end{table*}

\begin{table}
\caption{Measured Hyperfine coefficients $A$, $B$ of each UV transition.}
\label{table:3}
\begin{tabularx}{\columnwidth}{Y Y Y Y}
\hline\hline
 \makecell{Wavelength\\(nm)} & \makecell{Isotope\\} & \makecell{$A$ (MHz)\\} & \makecell{$B$ (MHz)\\} \\
 \hline
 359.05  & 161 & -128.4(18) & 1253(58) \\
          & 163 & 181.1(18) & 1316(58)\\
 359.26 & 161 & -113.2(21) & 475(59)\\ 
          & 163 & 160.9(21) & 471(59)\\
 359.48 & 161 & -121.2(18) & 1364(59)\\ 
          & 163 & 169.3(18) & 1480(59)\\
 362.89 & 161 & -90.1(21) & 1272(58)\\
          & 163 & 123.7(21) & 1383(58)\\
 372.19 & 161 & -78.7(17) & 1513(58)\\
          & 163 & 110.3(12) & 1595(57)\\
\hline\hline
\end{tabularx}
\end{table}

\begin{table}
\caption{Error budget for the \SI{362.89}{\nano m} transition.}
\label{table:4}
\begin{tabularx}{\columnwidth}{Y Y} 
 \hline\hline
 \text{Source} & \text{Uncertainty (MHz)} \\  
 \hline
 Wavemeter-scope signal alignment & max 1.1 \\
 Wavemeter accuracy  & 6.7\\ 
 Blue frequency error & 6.8 \\
 Elliptical 2D-Gaussian fitting error & $< $ 0.1\\ 

 \hline
 Total & max 9.6 \\
 \hline\hline
\end{tabularx}
\end{table}

In addition to the dominant resonances, 7 additional weaker resonances indicated with arrows in Figure \ref{fig:2}(a) are identified. These resonances also appear at the same $\Delta\nu_{421}$ as dominant hyperfine transitions. We thus attribute them to the less probable $\Delta F = 0$ excitations between ground and excited hyperfine states. Based on the measured hyperfine structure, their positions appear at frequencies $\Delta\nu_{\text{UV}}$ which coincide with transitions of this type, where they are observed at more (less) positive frequencies with respect to their associated $\Delta F = -1$ transition for \isotope[163]{Dy} (\isotope[161]{Dy}), a consequence of hyperfine states with larger $F$ being higher (lower) in energy for \isotope[163]{Dy} (\isotope[161]{Dy}). The resonances to $F' = 6.5$ and 7.5 for \isotope[163]{Dy} and $F' = 5.5$ for \isotope[161]{Dy} were outside of the used $\Delta\nu_{\text{UV}}$ scan range. The absence of such weak resonances at $\Delta\nu_{421}$ $\approx$ \SI{270}{\mega\hertz} (\SI{2170}{\mega\hertz}), even after scanning \SI{2}{\giga\hertz} above (below) the respective $\Delta F = -1$ resonance is in accordance with the absence of any $\Delta F = 0$ transition from the $F = 10$ hyperfine ground state.  

By extracting each resonance's UV position $\Delta\nu_{0,\text{UV}}$ and using the known ground state hyperfine coefficients for dysprosium \cite{ferch1974}, we determine the isotope shifts (Table~\ref{table:2}) and the hyperfine splittings and coefficients $A, B$ (Table~\ref{table:3}). Furthermore, we provide an accurate determination of the absolute wavenumber of each transition for the most abundant isotope \isotope[164]{Dy} (Table~\ref{table:1}). Previous measurements of isotope shifts for these transitions only provide values for $\delta\nu_{164-160}$ \cite{ahmad1982, afzal2000}. In general, we find reasonable agreement with the values stated, however for the transition \SI{362.89}{\nano m} (\SI{372.19}{\nano m}) where $\delta\nu_{164-160}$ is reported to be \SI{-660}{\mega\hertz} (+\SI{60}{\mega\hertz}), we measure significantly different values of \SI{-424(14)}{\mega\hertz} (\SI{-339(13)}{\mega\hertz}).

The quoted uncertainties for the values shown in Tables~\ref{table:2} and \ref{table:3} originate from the uncertainty on $\Delta\nu_{0,\text{UV}}$ of each resonance, which we briefly describe here (see supp. mat. for details). Table~\ref{table:4} shows the sources of error on the resonance positions for the \SI{362.89}{\nano m} transition, with similar values obtained for the other transitions studied. Firstly, alignment of the recorded lock-in output to the simultaneously recorded wavemeter data produces a maximum error of \SI{1.1}{\mega\hertz}, limited by the wavemeter sampling rate. The wavemeter used provides an expected absolute 3$\sigma$ accuracy of \SI{10}{\mega\hertz} on the recorded Ti:Sa laser frequency, resulting in a 1$\sigma$ accuracy of \SI{6.7}{\mega\hertz} for the UV frequency. Ideally, since each resonance is Doppler broadened by $\approx$ \SI{200}{\mega\hertz} in $\Delta\nu_{\text{UV}}$ due to residual atomic beam divergence, extraction of the $\Delta\nu_{0,\text{UV}}$ position of each resonance should correspond to the same velocity class before calculating isotope shifts for that transition. To ensure this consistency in our procedure, we fit a 2D elliptical Gaussian to every resonance, rotated to match the expected gradient based on the wavevector ratio $k_{421}/k_{\text{UV}}$. From the coordinates ($\Delta\nu_{0,\text{UV}}, \Delta\nu_{0,\text{421}}$) of the peak of the Gaussian fit, we compare each $\Delta\nu_{0,\text{421}}$ to the known isotope or hyperfine transition shifts relative to \isotope[164]{Dy} for the \SI{421}{\nano m} transition \cite{leefer2009}. We estimate a common standard blue frequency error $\delta(\Delta\nu_{0,\text{421}})$ from the statistical distribution of the differences, which translates to an error on $\Delta\nu_{0,\text{UV}}$ using $k_{421}/k_{\text{UV}}$. The fit error on the parameters ($\Delta\nu_{0,\text{UV}}, \Delta\nu_{0,\text{421}}$) is negligible compared to all other error sources, and the method of using the \SI{421}{\nano m} transition as a blue frequency error reference also captures any frequency drifts from the ULE cavity.  

\begin{figure}[t]
    \centering
    \includegraphics[]{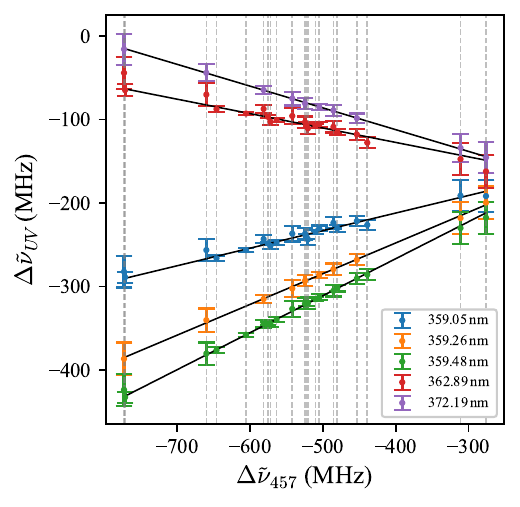}
    \caption{King plots. Vertical dashed lines indicate isotope shifts for the \SI{457}{\nano m} transition and correspond to all 21 isotope shift combinations between all stable Dy isotopes. Errors bars are calculated using the error budget (Table~\ref{table:4}) and are also normalized to the difference in nucleon number for a particular isotope shift combination.}
    \label{fig:3}
\end{figure}

We now discuss the electronic nature of each transition based on King plot analysis. From the extracted isotope shifts of each UV transition, we create a combined King plot for all transitions shown in Figure \ref{fig:3}. We plot the normalized isotope shifts $\Delta\nu_{\text{UV}}/\Delta N = \Delta\tilde\nu_{\text{UV}}$, where $\Delta N$ is the difference in isotope number, as a function of the normalized isotope shifts $\Delta\tilde\nu_{457}$ of a reference pure single-electron 4f$^{10}$6s$^{2}$ $\rightarrow$ 4f$^{10}$6s6p transition at 457\,nm \cite{zaal1980}. After performing a least-squares weighted linear fit to each data set, we extract the specific mass shifts (SMS) and electronic field shift (EFS) ratios $E_{\text{UV}}/E_{457}$ (Table~\ref{table:5}) from the King plot (see supp. mat.) Besides the normal mass shift which can be calculated analytically \cite{zaal1980}, the SMS arises from the influence of electronic momentum pair correlations, and the EFS arises from the varying overlapping charge density between the nucleus and electrons, which depends on the specific electronic configuration and nucleon number. Starting from the top of Figure \ref{fig:3}, the negative slopes given by the EFS ratios $E_{372}/E_{457}$ and $E_{362}/E_{457}$ for the
\SI{372.19}{\nano m} and \SI{362.89}{\nano m} transitions respectively highlights their significantly different electronic nature compared to the \SI{457}{\nano m} reference. EFS ratios of -0.261(0.002) and -0.173(0.012), as well as the relatively large negative SMSs of \SI{-475(3)}{\mega\hertz} and \SI{-433(13)}{\mega\hertz} respectively indicate their nature as pure two-electron transitions of type 4f$^{10}$6s$^{2}$ $\rightarrow$ 4f$^{9}$5d$^{2}$6s, where our measured values are in good agreement with reported values for such transitions \cite{zaal1980}. These values also confirm previously suggested configuration assignments \cite{wyart1974, li2017} and the accuracy of our measured $\delta\nu_{164-160}$ values for these two transitions.

The other three transitions we study are more complex in nature due to stronger configuration mixing. The closeness in energies, common odd parity and same $J$ between the pairs of levels corresponding to the \SI{359.05}{\nano m} and \SI{359.48}{\nano m} transitions, as well as between the \SI{359.26}{\nano m} and \SI{357.34}{\nano m} transitions is expected to lead to mixing \cite{ahmad1982, li2017}. For the \SI{359.05}{\nano m} transition which has a leading 4f$^{9}$5d$^{2}$6s configuration \cite{wyart1974}, mixing is highlighted from its relatively weak positive King plot slope of 0.210(0.010) (Figure \ref{fig:3}). Furthermore, we extract a SMS of \SI{-283(12)}{\mega\hertz}, $\approx$ \SI{190}{\mega\hertz} higher than for typical pure 4f$^{9}$5d$^{2}$6s transitions \cite{zaal1980}. This is consistent with significant mixing with the level corresponding to the \SI{359.48}{\nano m} transition, which has a leading 4f$^{10}$6s6p configuration \cite{wyart1974}. Typical SMS values for pure 4f$^{10}$6s6p transitions are $\approx$ \SI{8}{\mega\hertz} \cite{zaal1980, leefer2009, jin2011}. Consequently, we extract an influenced negative SMS of -198(5) MHz for the \SI{359.48}{\nano m} transition with a corresponding King plot slope of 0.443(0.004), in contrast to pure 4f$^{10}$6s6p transitions which would have a slope $\approx$1 \cite{leefer2009}. In a similar way, the level of the  \SI{359.26}{\nano m} transition of leading 4f$^{10}$6s6p configuration \cite{wyart1974} mixes with the level of the \SI{357.34}{\nano m} transition with leading 4f$^{9}$5d$^{2}$6s configuration \cite{wyart1974}, resulting in an extracted positive slope of 0.368(0.005) and SMS of \SI{-223(6)}{\mega\hertz}. 

\begin{table}
\caption{Measured SMS for $\delta_{164-162}$ and electronic field shift ratios.}
\label{table:5}
\begin{threeparttable}
\begin{tabularx}{\columnwidth}{Y Y Y Y}
\hline\hline
\makecell{Wavelength\\(nm)} & \makecell{SMS (MHz)\\} & \makecell{$\text{E}_\text{UV}/\text{E}_{\text{457}}$\\} &\makecell{Configuration\\}  \\
 \hline
 359.05\tnote{a}  & -283(12) & 0.210(0.010) & 4f$^{9}$5d$^{2}$6s + 4f$^{10}$6s6p\\
 359.26\tnote{b}  & -223(6) & 0.368(0.005) & 4f$^{10}$6s6p + 4f$^{9}$5d$^{2}$6s\\ 
 359.48\tnote{a} & -198(5) &  0.443(0.004) & 4f$^{10}$6s6p + 4f$^{9}$5d$^{2}$6s\\
 362.89 & -433(13) & -0.173(0.012) & 4f$^{9}$5d$^{2}$6s\\
 372.19 & -475(3) & -0.261(0.002) & 4f$^{9}$5d$^{2}$6s\\
\hline\hline
\end{tabularx}
\begin{tablenotes}
\item[a] mixed with each other
\item[b] mixed with a level corresponding to \SI{357.34}{nm} of configuration 4f$^{9}$5d$^{2}$6s ($^{9}$L$_{7}$) \cite{wyart1974, martin1978}
\end{tablenotes}
\end{threeparttable}
\end{table}

\begin{figure*}[ht]
    \centering
    \includegraphics[]{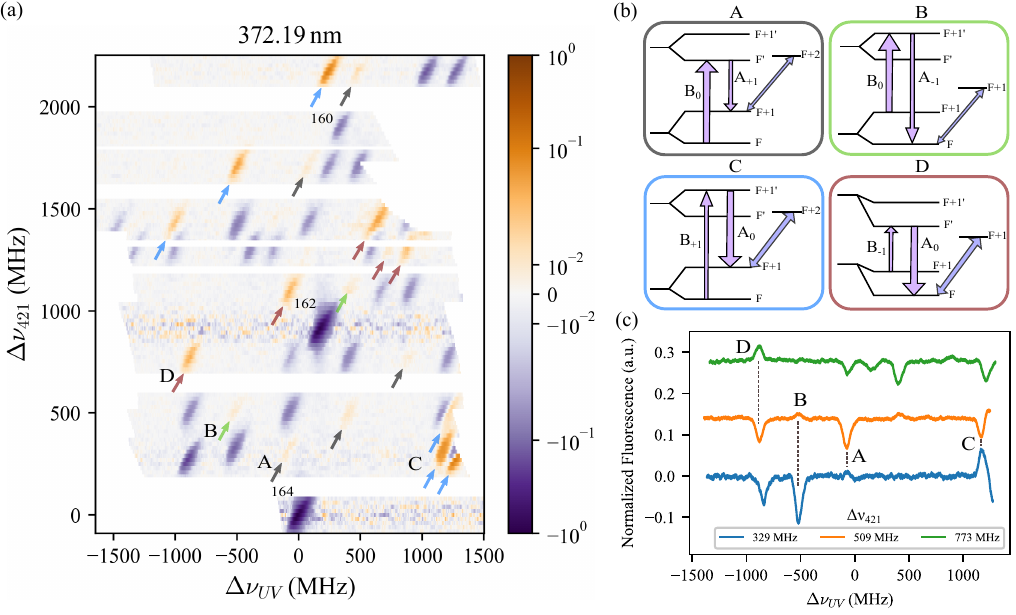}
    \caption{(a) Fluorescence map of the \SI{372.19}{\nano m} transition, normalized to the absolute value of the minimum corresponding to the most abundant isotope \isotope[164]{Dy}. Resonances for the bosons \isotope[164]{Dy}, \isotope[162]{Dy} and \isotope[160]{Dy} are labeled. \isotope[158]{Dy} and \isotope[156]{Dy} were not detected. White spaces indicate frequency intervals where we did not measure. The colorbar indicates the normalized fluorescence (a.u.), being linear between $\pm 0.035$, with a log scale applied outside this interval. (b) Simplified energy level schemes of the fluorescence resonances. Arrows colored to the corresponding resonance type are shown in (a). Weak resonances appear when the UV excitation is a dominant hyperfine transition ($\Delta F = 0$), where some atoms then weakly decay to a hyperfine ground state that is either above (A, $\Delta F = +1$) or below (B, $\Delta F = -1$) the initial ground hyperfine state. Strong resonances appear when the UV excitation is weak (C, $\Delta F = +1$ or D, $\Delta F = -1$) with strong decay occurring between hyperfine states with $\Delta F = 0$. (c) Individual traces (offset for clarity) of the fluorescence map at 3 different $\Delta\nu_{421}$ frequencies. Measured resonances types in (b) are indicated, with the fluorescence peak sharing the same $\Delta\nu_{\text{UV}}$ as a shelving resonance.}
    \label{fig:4}
\end{figure*}

From the 5 transitions we study in detail, only atoms excited on the \SI{372.19}{\nano m} transition have a significant relative decay strength back to the ground state. The branching ratio $A_{\text{GS}}/(A_{\text{FES}}+A_{\text{GS}})$ $\approx$ 3/4 (Table~\ref{table:1}) is much larger in comparison to the other transitions where the majority of atoms decay to the FES. Since the transition is also comparable in strength to the \SI{359.05}{\nano m} and \SI{362.89}{\nano m} transitions (Table~\ref{table:1}) which have a known average decay time back to the ground state of $\approx$ \SI{4.4}{\micro\second}, $\approx$ 3/4 of the atoms that are excited decay back to the ground state before reaching the blue beam. These atoms can then be excited by the blue beam on the strong \SI{421}{\nano m} transition. We observe the expected shelving resonances for bosons and fermions, as well as resonances of increased fluorescence, plotted in purple and orange respectively (Figure \ref{fig:4}(a)). These resonances appear only at $\Delta\nu_{\text{UV}}$ frequencies where a particular shelving resonance for a hyperfine transition exists, but have a different $\Delta\nu_{421}$ to the shelving resonance. Thus, we attribute them to atoms that were initially excited to a particular UV hyperfine state being optically pumped to different hyperfine ground states via spontaneous decay.

The fluorescence resonances appear weak (strong) when the initial UV excitation is strong, $\Delta F = 0$ (weak, $\Delta F = \pm 1$) and we illustrate the 4 possible scenarios in Figure \ref{fig:4}(b). Due to the inverted hyperfine structure between \isotope[161]{Dy} and \isotope[163]{Dy}, resonances of type `A' and `C' appear at $\Delta\nu_{421}$ frequencies that are lower (higher) than their corresponding shelving resonance for \isotope[163]{Dy} (\isotope[161]{Dy}), and vice versa for type `B' and `D'. By detecting both shelving and fluorescence resonances with the same blue fluorescence signal, we deduce the relative strength between the UV excitation and decay rates, which is typically not possible with single beam methods that measure either only in absorption or fluorescence. Using a simple rate equation model and comparing resonance strengths (Figure \ref{fig:4}(c)), we find we work in the regime where $B_{+/-1}/B_{0} > A_{+/-1}/A_{0}$, where $B_{+/-1}$ ($B_{0}$) is the weak (strong) UV excitation rate and $A_{+/-1}$ ($A_{0}$) is the weak (strong) decay rate to the hyperfine state resonant with the blue beam. In a similar way, by evaluating the ratios between the magnitudes of a shelving resonance and measured fluorescence resonances that share the same $\Delta\nu_{\text{UV}}$, this directly comparable signal of the excitation and decay population rates can be used as a convenient method for evaluating the presence and strength of dark-state decay channels in a single measurement (see supp. mat.).

With our employed shelving technique, we show how $J$ of the excited state can be determined without any fitting of the measured spectra, prior knowledge of the electronic configuration, varying light polarization or applied magnetic field. From the orange trace in Figure \ref{fig:4}(c) which is measured when the blue beam is on resonance with the \isotope[163]{Dy} $F = 8.5$ hyperfine ground state ($\Delta\nu_{421}$  $\approx$ \SI{509}{\mega\hertz}), we measure one dominant shelving resonance at $\Delta\nu_{\text{UV}}$ $\approx$ \SI{-75}{\mega\hertz} and 2 weaker resonances at $\approx$ \SI{-875}{\mega\hertz} and \SI{1160}{\mega\hertz}. As $J = 8$ for the ground state, detection of a weaker resonance at a frequency lower than the dominant resonance at \SI{-75}{\mega\hertz} excludes $J = 7$ for the excited state, since $\Delta F = -2$ is forbidden. Similarly, the weaker resonance at a higher frequency excludes $J = 9$ as $\Delta F = +2$ is forbidden \cite{hyperfineComment}. Hence, the only remaining possibility is $J = 8$ which is in agreement with the assigned value from Ref. \cite{conway1971}. We note that the relative strengths of the resonances corresponding to the three transitions ($\Delta F = 0, \pm 1$) can vary depending on the population distribution between the ground state $m_{F}$ sub-levels. Indeed, we measure a lower relative amplitude ratio between the dominant and weaker resonances compared to strengths based on a uniform population distribution (supp. mat. Table~\ref{table:S2}), possibly due to a stray magnetic field in our chamber influencing the distribution via the Hanle effect. Despite the seemingly non-uniform distribution in our experiment, this technique remains robust to some atomic redistribution of $m_{F}$ states, since the $\Delta F = \Delta J$ resonance is always significantly stronger than the other two weaker resonances for a uniform distribution (supp. mat. Table~\ref{table:S2}). For future measurements of transitions where $J$ is not known, the $\Delta F = \Delta J$ resonance is always measured as the dominant resonance by either working with a uniform $m_{F}$ distribution or exciting with isotropic (equal linear and circular-polarized components) light. Thus, this straightforward extraction of $J$ works with any observed shelving triplet from any hyperfine ground state.

Finally, we want to comment on the 5 transitions reported previously (supp. mat., Table \ref{table:S1}) which were not observed in this work. For each of these transitions, we performed an extensive search by first fixing $\Delta\nu_{421}$ to be resonant with the most abundant isotope \isotope[164]{Dy}, then scanned $\approx$ \SI{2}{\giga\hertz} around the expected UV resonance frequency. No signals above the noise level were observed this way. While four of the five transitions are expected to be weak and are potentially below our detection limit, the \SI{370.18}{\nano m} transition (27014 cm$^{-1}$) has been reported to be of similar strength as the strong \SI{421}{\nano m} transition \cite{wickliffe2000, curry1997}. As speculated previously in theoretical work \cite{li2017}, this transition was likely incorrectly assigned as a ground state transition, consistent with our measurement. 

\section{Conclusion}
We have measured the isotope shifts, hyperfine structure and determined the electronic nature of multiple UV states accessible from the ground state that are ideal for optically populating the FES in Dy. These values provide the necessary information to either efficiently optically pump or perform fast coherent population transfer to the FES with a two-photon Raman transition, where the coupling wavelength from the UV state to the FES is conveniently close to the commonly used \SI{421}{\nano m} light in Dy experiments. We expect these UV states which analogously exist in other magnetic lanthanides to enable increased access to the FES in various ultracold atom experiments. Furthermore, we have demonstrated the use of two-dimensional shelving spectroscopy that exploits the strong \SI{421}{\nano m} transition to enhance detection sensitivity and greatly simplify the extraction of all relevant quantities. This technique is therefore also practical for characterizing any atomic transition in conjunction with typical wavelengths used for cooling and imaging.

\begin{acknowledgments}
We thank the QRydDemo team in Stuttgart for allowing us to use their wavemeter for our measurements. We acknowledge J.-N. Schmidt and V. Anasuri for preliminary measurements on two of the transitions studied in this work. This work was funded by the European Research Council (grant agreement No. 101019739).  J.H. gratefully acknowledges support by the MIT Pappalardo Fellowships in Physics. S.W. acknowledges support from the Center for Integrated Quantum Science and Technology (IQST) and financial support from the German Research Foundation through the Emmy Noether Grant No. WE 7554/1-1, and the Carl-Zeiss-Stiftung Center for Quantum Photonics (QPhoton).
\end{acknowledgments}

\bibliography{references}


\iftoggle{arXiv}{
	\clearpage 
	\title{Supplementary Material: Two-dimensional shelving spectroscopy of ultraviolet ground state transitions in dysprosium}
    \maketitle
}{}
\label{supplement}
\section*{UV transitions}
Table \ref{table:S1} shows the electric-dipole allowed UV ground state  transitions between \SI{359.0}{\nano m} and \SI{372.5}{\nano m} investigated in this work.

\renewcommand{\thetable}{S1}
\begin{table*}[ht]
\caption{Reported \cite{conway1971, martin1978, wyart1974, wickliffe2000} electric-dipole allowed ($\Delta J$ = 0,$\pm$1) ground state UV transitions between 359.0 - \SI{372.5}{\nano m} in neutral dysprosium investigated in this work. If no leading term symbol or configuration is reported in the references, only $J$ is written.}

\label{table:S1}
\begin{tabularx}\textwidth{Y Y Y Y Y} 
\hline\hline
\text{Wavelength (nm)} & \text{Wavenumber (cm$^{-1}$)} & \makecell{Leading term\\ symbol ($^{2S+1}L_{J}$)} & \text{Leading configuration} & \text{Detected in this work?}\\  
\hline
 359.05 & 27851 & $^{7}$I$_{8}$ & 4f$^{9}$5d$^{2}$6s & yes \\ 
 359.23 & 27838 & $J = 7$  & - & no\\
 359.26 & 27835 & $^{5}$I$_{7}$ & 4f$^{10}$6s6p &  yes\\ 
 359.48 & 27818 & $^{5}$K$_{8}$ & 4f$^{10}$6s6p & yes\\
 362.89 & 27556 & $^{9}$L$_{7}$ & 4f$^{9}$5d$^{2}$6s & yes\\ 
 364.60 & 27427 & $^{5}$K$_{7}$ & 4f$^{10}$6s6p & no \\
 366.04 & 27319 & $J = 8$  & - & no\\
 366.08 & 27316 & $^{7}$G$_{7}$ & 4f$^{9}$5d$^{2}$6s & yes\\
 370.18 & 27014 & $J = 9$  & - & no\\
 372.19 & 26868 & $J = 8$  & - & yes\\
 372.49 & 26846 & $J = 7$  & - & no\\
 \hline\hline
\end{tabularx}
\end{table*}

\section*{Shelving enhancement factor}

Using the \SI{359.05}{\nano m} and \SI{362.89}{\nano m} transitions where the branching ratios are known, the shelving enhancement factor is estimated by calculating the decrease in number of scattered \SI{421}{\nano m} photons due to a single UV photon excitation.  

In our experiment for these two transitions, shelving of atoms to the UV state which decay directly back to the ground state will not contribute to any shelving signal, as the travel time between the UV and blue beams is longer that the direct decay time back to the ground state. Thus, only atoms that decay to the FES cause \SI{421}{\nano m} photon loss. Also, since the FES lifetime is significantly longer than both the interaction time between atoms and the blue beam $\tau_{\text{blue}}$, as well as the travel time between the UV and the blue beams, the measured photon loss is limited by $\tau_{\text{blue}}$. 

For atoms traveling at $\approx$ \SI{500}{ m/\second} passing through the UV beam with waist $\approx$ \SI{0.47}{\milli m}, the interaction time between the atoms and the UV beam $\tau_{\text{UV}}$ $\approx$ \SI{1.88}{\micro\second} is less than the average decay time back to the ground state of \SI{4.4}{\micro\second}. Thus on average, less than one UV photon is scattered. Also, for a blue beam with waist $\approx$ \SI{0.85}{\milli m}, $\tau_{\text{blue}}$ $\approx$ \SI{3.4}{\micro\second}. Using our blue beam parameters, the decreased number of scattered \SI{421}{\nano m} photons is thus 

\begin{align*}
R_{\text{loss,blue}} &= \frac{A_{\text{FES}}}{A_{\text{FES}} + A_{\text{GS}}}\frac{A_{421}}{2}\frac{s}{1+s}\cdot \tau_{\text{blue}} \\
& = 0.93\cdot\frac{2.02 (10)^{8} s^{-1}}{2} \frac{1.3}{1+1.3}\cdot 3.4\, \text{\textmu s}\\ 
&\approx181. 
\end{align*}
where $A_{\text{FES}}/(A_{\text{FES}} + A_{\text{GS}})$ is the average branching ratio to the FES for the two UV states considered, $A_{421}$ is the $A$ coefficient of the \SI{421}{\nano m} transition and $s$ is the blue beam saturation parameter.   

\section*{Error evaluation} 

The procedure for creating spectroscopic maps such as the one in Figure \ref{fig:2}(a), the extraction of errors in Tables~\ref{table:2},~\ref{table:3} and \ref{table:5} and the error budget in Table~\ref{table:4} is described below. 

\subsection{Spectroscopic map creation}

At each $\Delta\nu_{421}$, we ramp the UV frequency $\Delta\nu_{UV}$ controlled by the sawtooth voltage output of the laser that varies the Ti:Sa light frequency via an intra-Ti:Sa cavity piezo mirror. Each ramp ends after \SI{4500}{\milli\second} and is repeated eight times. The faster voltage ramp-back that occurs before the next ramp takes approximately \SI{20}{\milli\second}. Simultaneously for each ramp, we record the lock-in amplifier (LIA) voltage and the wavemeter frequency output. Both the laser control voltage and the LIA voltage are recorded on the same oscilloscope and share the same time base vector. The sampling rates of the recorded signals on the oscilloscope and on the wavemeter are around $10^5$ and $200$ samples per second, respectively. 

To align the recorded frequency for each ramp to the LIA output, we first align the frequency data to the laser control voltage ramp. We perform a peak-finding routine to identify the last recorded frequency value associated with the end of the ramp control voltage. A two-segment line is then fit to points around the found peak and we take the extracted time point at the intersection between the line segments as an estimate of the time corresponding to the endpoint of the laser ramp control voltage. This intersection time is then subtracted from the respective frequency data time values to align the recorded frequency values to the laser control voltage. Following this, we perform a 9$^{\text{th}}$ order polynomial fit of the frequency which captures systematic scanning nonlinearities originating from the piezo-mounted Ti:Sa cavity mirror. The fit error after averaging across the eight ramps (see below) is negligible compared to the other dominant error sources. The fit function is then called at each time point defined by the time base vector from the oscilloscope. As a result, the extracted frequency values then correspond to the same times as the LIA output. 

After doubling the aligned Ti:Sa frequencies and taking into account the \SI{110}{\mega\hertz} shift from the UV AOM to obtain the absolute scanned UV frequency, we obtain a trace of the LIA output as a function of UV frequency for each ramp. We then interpolate each trace at a defined frequency vector with a resolution that approximately matches the resolution of the oscilloscope data. This additional interpolation step is performed so that the respective points of each trace correspond to the same frequency value. Finally, we average the LIA output across the eight traces at each frequency value before creating the spectroscopic map. The frequency axis is set relative to the extracted \isotope[164]{Dy} resonance position ($\Delta\nu_{0,\text{UV}}, \Delta\nu_{0,\text{421}}$) from the 2D Gaussian fit. For transitions where we measure \isotope[158]{Dy} and \isotope[156]{Dy}, the gain of the PMT for these traces was increased with respect to the other traces of the map. Hence, we scale the traces of \isotope[158]{Dy} and \isotope[156]{Dy} by a common factor using the known natural abundance of these isotopes with respect to \isotope[164]{Dy}. Any signal offset between the traces for \isotope[158]{Dy} and \isotope[156]{Dy} relative to other traces of the map were also accounted for to obtain a common signal background. The spectroscopic maps for the \SI{359.05}{\nano m}, \SI{359.26}{\nano m} and \SI{359.47}{\nano m} transitions are shown in Figures \ref{fig:S3}, \ref{fig:S4}, \ref{fig:S5}, respectively.

\subsection{Wavemeter-scope signal alignment}
The accuracy of using the intersection of the two-segment line as the reference point in time to align the recorded frequency data to the laser control voltage is limited by the wavemeter sampling rate. The value quoted in Table~\ref{table:4} is a conservative value for the maximum error possible due to misalignment. This is calculated by multiplying the nominal sampling rate of \SI{5}{\milli\second} by the scan speed of $\approx$ \SI{0.3}{\mega\hertz}/$\text{ms}$. Since the LIA voltage at each $\Delta\nu_{\text{UV}}$ is averaged over eight traces after alignment and interpolation, the final alignment error is further reduced by a factor of $\sqrt{8}$. We note that this error is not Gaussian-distributed. However, as the overall contribution of this error source is small and overestimated, the total error shown in Table~\ref{table:4} which is obtained by simply adding the error sources in quadrature is also conservative.

\subsection{Wavemeter accuracy}
Using the High Finesse wavemeter WS/8-2 approximately \SI{22}{\nano m} away from the wavelength used for calibration, the device provides an expected $3\sigma$ uncertainty of \SI{10}{\mega\hertz} for our Ti:Sa light which is frequency-doubled to the UV. Thus, a $1\sigma$ uncertainty for our UV frequency corresponds to \SI{6.7}{\mega\hertz}.

\subsection{Blue frequency error}
As described in the main text, errors in extracting the blue frequency resonance positions $\Delta\nu_{0,\text{421}}$ that do not correspond to the same velocity class produce an error on each extracted UV frequency resonance position $\Delta\nu_{0,\text{UV}}$. We calculate and apply a common nominal standard error $\delta(\Delta\nu_{0,\text{421}})$ from all resonances across the five transitions measured, since the accuracy of scanning the blue frequency was the same throughout the measurements and the extraction of all resonance positions were not limited by noise. 

To estimate $\delta(\Delta\nu_{0,\text{421}})$, we first calculate the distribution of the deviations between the expected and extracted values of $\Delta\nu_{0,\text{421}}$ for each resonance relative to \isotope[164]{Dy}, after setting the extracted blue frequency resonance position of \isotope[164]{Dy} resonances to \SI{0}{\mega\hertz}. The expected frequency shifts and associated errors of all bosonic isotopes and fermionic hyperfine resonances with respect to \isotope[164]{Dy} are calculated using the reported values of the isotope shifts and hyperfine coefficients of the \SI{421}{\nano m} excited state \cite{leefer2009} and ground state \cite{ferch1974}. From 75 resonances with extracted values of $\Delta\nu_{0,\text{421}}$ with respect to \isotope[164]{Dy}, we assign to each a common error $\delta(\Delta\nu_{0,\text{421}})$ and then use the relation $\sigma_{\bar{x}} = \sigma/\sqrt{N}$ between the standard deviation of the calculated distribution $\sigma$ and standard error on the mean of the distribution $\sigma_{\bar{x}}$, where $N$ is the number of samples. For our distribution where $\sigma_{\bar{x}} = (1/75)\sqrt{985.35 + 75 \delta(\Delta\nu_{0,\text{421}})^2}\text{MHz}$  and $\sigma =$ \SI{6.9}{\mega\hertz}, $\delta(\Delta\nu_{0,\text{421}}) = $ \SI{5.9}{\mega\hertz}. This value translates to a \SI{6.8}{\mega\hertz} error on each extracted $\Delta\nu_{0,\text{UV}}$ for the \SI{362.89}{\nano m} transition, varying slightly for the resonances corresponding to different UV transitions according to the wavevector mismatch $k_{421}/k_{\text{UV}}$.

For the error values given in Table~\ref{table:1} for the absolute value of the wavenumber of each transition for \isotope[164]{Dy}, the dominant error source in our measurement is systematic and originates from the uncertainty in extracting the resonance position corresponding to the zero-velocity class. Noting that lineshapes of traces measured exhibited some asymmetry, we conservatively estimate this error value by finding a blue frequency interval around the extracted resonance position from the 2D Gaussian fit which would contain a trace with the largest signal as well as a trace with a symmetric lineshape. This interval is possible to find due to the consistent behavior of the asymmetry and is $\pm$\SI{40}{\mega\hertz}, limited by our resolution in the blue frequency. This translates to $\pm$\SI{47}{\mega\hertz} in the UV frequency, again varying slightly between transitions due to $k_{421}/k_{\text{UV}}$. Furthermore, due to possible misalignment of the UV beam being not perfectly orthogonal to the atomic beam, an additional systematic error exists from the Doppler shift $k_{\text{UV}} v\cdot \text{cos}(90\degree-\theta)$, where $v$ is the atomic velocity of $\approx$ \SI{500}{ m/\second} and $\theta$ is the angle deviation from the UV beam being perfectly orthogonal to the zero-velocity class of the atomic beam. For a distance of $\approx $ \SI{178}{\milli m} used to align the UV beam with waist $\approx$ \SI{0.47}{\milli m} through the chamber, $\theta \approx 0.15\degree$ which results in a frequency shift of $\approx$ \SI{3.7}{\mega\hertz}. These two error sources are then added in quadrature to give the error values shown in Table~\ref{table:1}. 

\subsection{Elliptical 2D-Gaussian fitting error}
From the spectroscopic map, each identified resonance is fit with an elliptical, rotated two-dimensional Gaussian function \eqref{eq:twoDGauss}

\begin{widetext}
\begin{equation}
\begin{aligned}
f_{\text{gauss, 2D}}(\Delta\nu_{\text{UV}}, \Delta\nu_{421}) = 
    A\cdot \text{exp}\bigg[-\bigg(\frac{[(\Delta\nu_{\text{UV}} - \Delta\nu_{0,\text{UV}})\cdot \text{cos}(\theta) + (\Delta\nu_{421} - \Delta\nu_{0,\text{421}})\cdot \text{sin}(\theta)]^2}{2\sigma_{x}^2} \\ 
    &\hspace{-7.75 cm} + \frac{[-(\Delta\nu_{\text{UV}} - \Delta\nu_{0,\text{UV}})\cdot \text{sin}(\theta) + (\Delta\nu_{421} - \Delta\nu_{0,\text{421}})\cdot \text{cos}(\theta)]^2}{2\sigma_{y}^2}\bigg)\bigg] + f_{0}
\end{aligned}
\label{eq:twoDGauss}
\end{equation}
\end{widetext} where $A$ is the amplitude, $\theta = \text{arctan}(\lambda_{\text{UV}}/\lambda_{421})$ with $\lambda_{\text{U}V}$ and $\lambda_{421}$ being the wavelengths of the UV and \SI{421}{\nano m} transition respectively, $\Delta\nu_{0,\text{UV}}$, $\Delta\nu_{0,\text{421}}$ are the $x$ and $y$ coordinates of the peak position respectively, $\sigma_x$, $\sigma_y$ are the $x$ and $y$ Gaussian spread parameters and $f_{0}$ is some signal offset. The 1$\sigma$ statistical fitting error for the position of each resonance taken as the fit parameters ($\Delta\nu_{0,\text{UV}}$, $\Delta\nu_{0,\text{421}}$) is < \SI{0.1}{\mega\hertz}. Typical values for $\sigma_{x}$ are $\approx$ \SI{55}{\mega\hertz}, in reasonable agreement with the expected residual Doppler broadening due to atomic beam divergence which is limited by the aperture diameter. Typical values for $\sigma_y$ are $\approx$ \SI{30}{\mega\hertz}, in reasonable agreement with the moderately power-broadened linewidth of the \SI{421}{\nano m} transition. As mentioned above, lineshapes of individual resonance traces exhibited some asymmetric behavior which was consistent across all measured resonances. While it is difficult to isolate the cause of the asymmetry, any error caused by asymmetric lineshapes resulting in extracted resonance positions ($\Delta\nu_{0,\text{UV}}, \Delta\nu_{0,\text{421}}$) not corresponding to the same velocity class are accounted for with our method described in subsection D.

\renewcommand{\thetable}{S2}
\begin{table*}[ht]
\caption{Calculated hyperfine transition strengths for transitions from the ground state of Dy with $J = 8, I = 5/2, F = |J-I|,....,|J+I|$ to excited states where $\Delta J =0, \pm1$, $\Delta F = 0, \pm1$. The strongest transition is when $\Delta F = \Delta J$.}
\label{table:S2}
\begin{tabularx}{\textwidth}{Y|YYY|YYY|YYY}
\hline\hline
& \multicolumn{3}{c|}{$\Delta J = -1$} 
& \multicolumn{3}{c|}{$\Delta J = 0$} 
& \multicolumn{3}{c}{$\Delta J = +1$} \\
$F$ 
& $\Delta F = -1$ & $\Delta F = 0$ & $\Delta F = +1$ 
& $\Delta F = -1$ & $\Delta F = 0$ & $\Delta F = +1$ 
& $\Delta F = -1$ & $\Delta F = 0$ & $\Delta F = +1$ \\
\hline
5.5  & 0.944 & 0.054 & 0.001 &    - & 0.952 & 0.048        & - & - & 1 \\
6.5  & 0.925 & 0.074 & 0.001 &    0.041 & 0.893 & 0.065    & - & 0.037 & 0.963  \\
7.5  & 0.926 & 0.073 & 0.001 &    0.057 & 0.878 & 0.064    & 0.000 & 0.051 & 0.948 \\
8.5  & 0.942 & 0.058 & 0.000 &    0.057 & 0.892 & 0.051    & 0.001 & 0.051 & 0.948 \\
9.5  & 0.967 & 0.033 & - &    0.046 & 0.925 & 0.029        & 0.001 & 0.041 & 0.958 \\
10.5 & 1 & - & - &    0.026 & 0.974 & -                    & 0.000 & 0.024 & 0.976 \\
\hline\hline
\end{tabularx}
\end{table*}

\section*{Hyperfine Coefficients}
With each UV resonance position $\Delta\nu_{0,\text{UV}}$ given an error following the error budget of Table~\ref{table:4}, the hyperfine coefficients and errors in Table~\ref{table:3} are evaluated by first calculating the energetic ordering and frequency splittings between hyperfine states based on the extracted UV resonance frequencies and the known ground state hyperfine splittings \cite{ferch1974}. Using the equation for the splitting of a particular hyperfine state $F$ from the degenerate state 

\begin{align*}
\Delta E_{\text{hfs}} = AK + B\frac{(3/2)K(2K + 1) - I(I+1)J(J+1)}{2I(2I-1)J(2J-1)}
\end{align*}
where $ K = (1/2)(F(F+1) - J(J+1) - I(I+1))$, we then solve for $A$ and $B$ using pairs of splittings between three consecutive hyperfine states.
With this method of evaluation, we solve for $A$ and $B$ four times for transitions where all six possible hyperfine resonances in the manifold were measured. For the \SI{372.19}{\nano m} transition where we did not scan $\Delta\nu_{\text{UV}}$ far enough to detect the $F = 5.5$ state of \isotope[163]{Dy}, we solve for $A_{163}$ and $B_{163}$ three times. The values obtained for $A$ and $B$ were then averaged to give the values shown in Table~\ref{table:3}.

\section*{King plot analysis}
Isotope shifts only involving bosons are calculated by subtracting the measured transition frequencies. The isotope shifts $\delta\nu_{164-163}$ and $\delta\nu_{164-161}$ are calculated using a particular measured hyperfine transition frequency between ground and excited hyperfine states $F$ and $F'$, the splitting $\Delta E_{\text{hfs}}$ of $F$ ($F'$) from the degenerate state calculated using the known ground state \cite{ferch1974} (extracted) hyperfine coefficients, and the measured transition frequency for \isotope[164]{Dy}. We average all possible solutions obtained using measured hyperfine transitions to give the values shown in Table~\ref{table:2}. The errors for all isotope shifts are calculated using standard error propagation with errors added in quadrature. We assume that the error sources are uncorrelated since it is difficult to reliably measure the correlation between error sources in Table~\ref{table:4}. As a result, errors we present are conservative estimates.  

With the isotope shift values in Table~\ref{table:2}, we create the combined King plot in Figure \ref{fig:3} using the isotope shift values of the \SI{457}{\nano m} transition from \cite{zaal1980}. From a weighted linear fit of the normalized isotope shifts, we extract the specific mass shift and electronic field shift ratios in Table~\ref{table:5} using the relation between the normalized isotope shifts of two different transitions $i, j$ \cite{kingBook}

\begin{align*}
\delta\tilde\nu_j^{AA'} = \frac{E_j}{E_i} \delta\tilde\nu_i^{AA'} + \frac{\left(M_j - M_i(E_j/E_i)\right)}{AA'}
\end{align*}
where $A, A'$ refer to different isotope mass numbers, $M_i = M_{i, \text{nms}} + M_{i, \text{sms}}$ is the sum of normal and specific mass shifts, $M_{i, \text{nms}} = \nu_i/1836 $ where $\nu_i$ is the transition frequency, $M_{457\text{nm}, \text{sms}}$ = 7(8) MHz \cite{zaal1980} and $E_i$ is the electronic field shift parameter.

\section*{Hyperfine transition strengths}
The strength $S_{FF'}$ of a particular transition between hyperfine states $F \rightarrow F'$ for equal populations in all $m_{F}$ sub-levels or for an isotropic pump field is characterized by the sum of the strengths of transitions from any particular sub-level $\ket{F \ m_{F}}$ to all possible excited state sub-levels $\ket{F' \ m_{F}'}$ for a particular $F'$. Table~\ref{table:S2} shows the calculated strengths of hyperfine transitions $\Delta F = 0, \pm 1$ using equation \eqref{eq:transition_strengths} from all hyperfine ground states of Dy.

\begin{equation}
\begin{aligned}
S_{FF'} &= \sum_q (2F'+1)(2J + 1)\cdot\\
&\hspace{1cm}\begin{Bmatrix} J & J' & 1 \\ F' & F & I \end{Bmatrix}^2  |\bra{F\ m_{F} \ket{F'\ 1\ (m_{F}-q)\ q}}^2 \\
&= (2F'+1)(2J+1) \begin{Bmatrix} J & J' & 1 \\ F' & F & I \end{Bmatrix}^2
\label{eq:transition_strengths}
\end{aligned}
\end{equation}
where $m_{F} = m_{F}' + q$ and $q = 0, \pm 1$.


\section*{Rate equation model}
Here, we show the derivation of the relative excitation and decay regime described in the main text as well as the expression for the ratios of fluorescence and shelving signals. 

\renewcommand{\thefigure}{S1}
\begin{figure}[ht]
    \centering
    \includegraphics[]{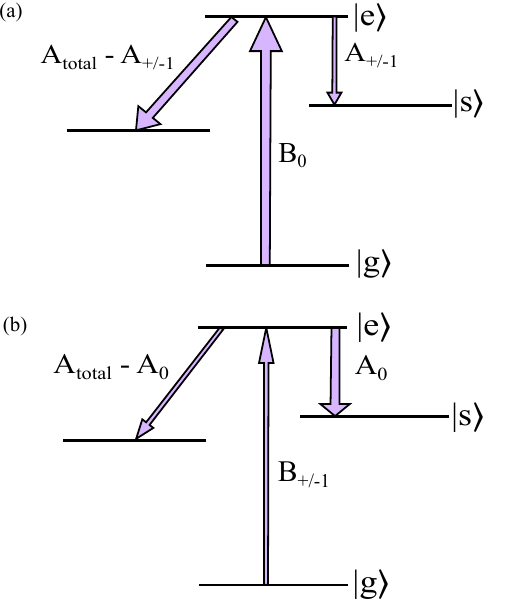}
    \caption{Energy level schemes of the (a) strong UV excitation $B_0$ and weak decay to the side hyperfine state $A_{+/-1}$ and (b) weak UV excitation $B_{+/-1}$ and strong decay to the side hyperfine state $A_0$ which result in weak and strong fluorescence signals, respectively.}
    \label{fig:S1}
\end{figure}

Figure \ref{fig:S1}(a)((b)) shows the relevant excitation and decay rates that describe the fluorescence resonances of type `A' and `B' (`C' and `D') in Figure \ref{fig:4}(b). $\ket{g}$, $\ket{e}$ and $\ket{s}$ denote the initial ground, excited UV hyperfine and side ground hyperfine state respectively. $B_0$ is the strong UV excitation rate, $B_{+/-1}$ is the weak UV excitation rate, $A_0$ is the strong decay rate to a hyperfine state, $A_{+/-1}$ is the weak decay to a hyperfine state, and $A_{\text{total}}$ is the total decay rate from $\ket{e}$. For Figure \ref{fig:S1}(a) ((b)), $A_{\text{total}} - A_{+/-1} (A_0)$ is the decay to all other possible channels. 

For the scheme of Figure \ref{fig:S1}(a), the atomic population in the side hyperfine state $\ket{s}$ which is directly proportional to the strength of the measured fluorescence resonance is $N_{s, a} \simeq N_g B_0 \tau_{\text{UV}}(A_{+/-1}/A_\text{total})$ where $N_g$ is the number of atoms initially in the ground state. In deriving this integrated rate equation, we have assumed a sufficiently short interaction time with the UV light such that multi-photon scattering is negligible, as well as a long enough travel time between UV and blue beams, which allows us to fully integrate out the (UV) excited state.   
Similarly for the scheme of Figure \ref{fig:S1}(b), $N_{s, b} \simeq N_g B_{+/-1}\tau_{\text{UV}}(A_{0}/A_\text{total})$. We observe that $N_{s, b} > N_{s, a}$ and thus $B_{+/-1}/B_0 > A_{+/-1}/A_0$.

\renewcommand{\thefigure}{S2}
\begin{figure}[ht]
    \centering
    \includegraphics[]{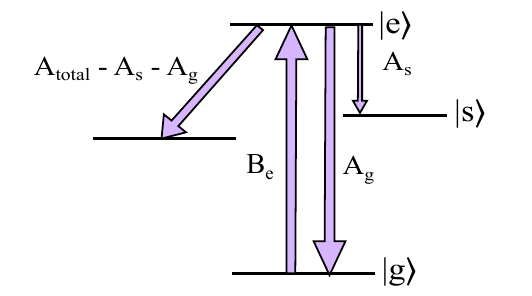}
    \caption{General energy level scheme of shelving with the UV excitation rate $B_e$, decay rate back to the ground state $A_g$, to a side hyperfine state $A_\text{s}$ and other remaining channels $A_{\text{total}} - A_\text{s} - A_\text{g}$.}
    \label{fig:S2}
\end{figure}

Figure \ref{fig:S2} shows the general scheme in which the relevant excitation and decay rates which determine the strength of a shelving resonance are labeled. Here, $B_\text{e}$ is the UV excitation rate, $A_\text{g}$ is the decay rate from $\ket{e}$ back to the ground state, $A_\text{s}$ is the decay to some side hyperfine state and $A_{\text{total}} - A_\text{s} - A_\text{g}$ is the decay to all other possible channels. For the \SI{359.05}{\nano m}, \SI{362.89}{\nano m} and \SI{372.19}{\nano m} transitions, atoms that decay directly back to the ground state after being excited do not contribute to any shelving signal, since this decay time is shorter than the travel time between UV and blue beams. Thus, the shelving signal originates only from atoms that decay to other states and is directly proportional to $N_g B_\text{e} \tau_{\text{UV}}(A_{\text{total}}-A_\text{g})/A_\text{total}$. As described similarly above, any fluorescence signal is proportional to $N_g B_\text{e}\tau_{\text{UV}}(A_\text{s}/A_\text{total})$ and the ratio between the magnitudes of a fluorescence and shelving resonance that share the same $\Delta\nu_{\text{UV}}$ is therefore $A_\text{s}/(A_{\text{total}}-A_\text{g})$, which is maximum 1 when decay to $\ket{s}$ is the only other existing side decay channel. As a result, when the ratio is less than 1, other decay channels must exist. 

\clearpage          
\begingroup
\setlength{\topskip}{0pt}  
\setlength{\parskip}{0pt}

\onecolumngrid      
\section*{Additional spectroscopic maps}
\endgroup

\renewcommand{\thefigure}{S3}
\begin{figure}[ht]
    \centering
    \includegraphics[]{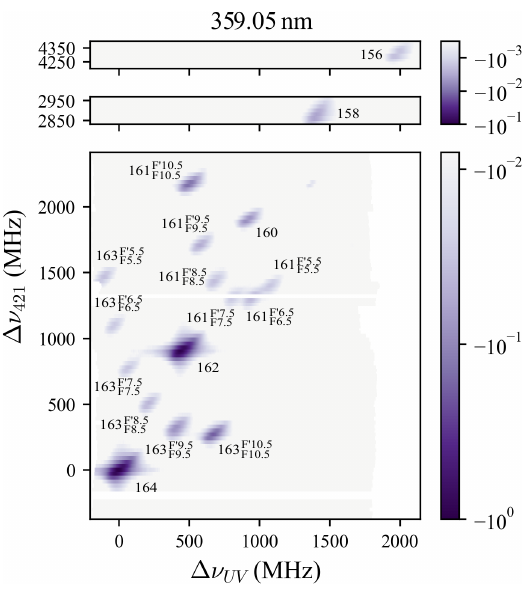}
    \caption{\SI{359.05}{\nano m} transition.}
    \label{fig:S3}
\end{figure}

\renewcommand{\thefigure}{S4}
\begin{figure}[ht]
    \centering
    \includegraphics[]{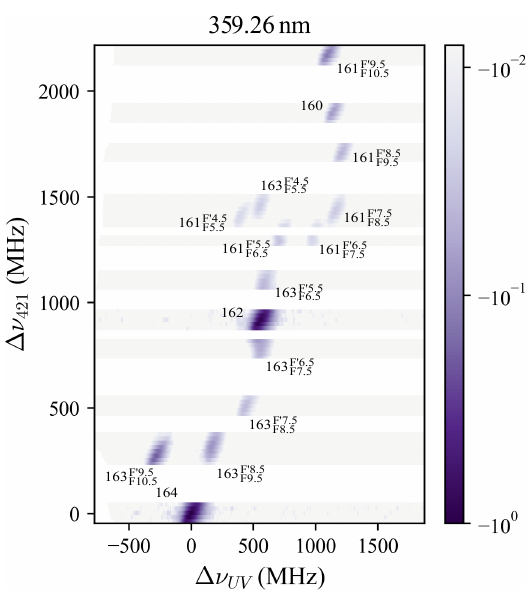}
    \caption{\SI{359.26}{\nano m} transition.}
    \label{fig:S4}
\end{figure}

\renewcommand{\thefigure}{S5}
\begin{figure}[ht]
    \centering
    \includegraphics[]{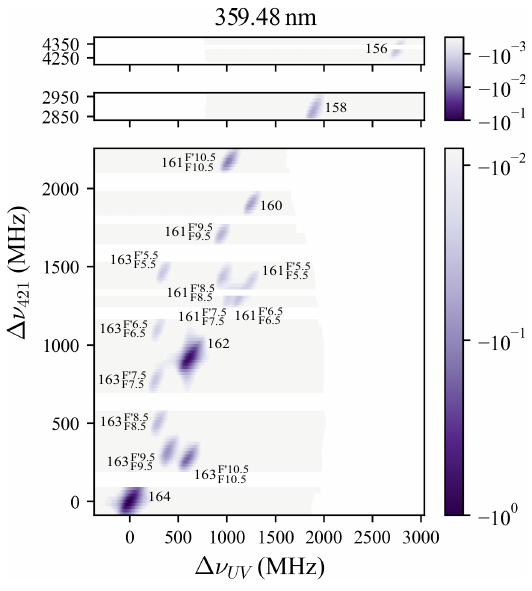}
    \caption{\SI{359.48}{\nano m} transition.}
    \label{fig:S5}
\end{figure}
\twocolumngrid
\end{document}